\documentclass[a4paper,fleqn,usenatbib,useAMS]{mnras}
\usepackage{graphicx}
\usepackage{amsmath}
\usepackage{dcolumn}
\usepackage{color}
\usepackage{epstopdf}
\usepackage[caption=false]{subfig}

\newcommand\lb{\langle}
\newcommand\rb{\rangle}
\newcommand\bOmega{\overline\Omega}
\newcommand\emfb{\overline{\mbox{\boldmath ${\cal E}$}} {}}
\newcommand\emf{\overline{\mbox{${\cal E}$}} {}}

\newcommand\bbB{\overline{\bf B}}
\newcommand\bV{\overline V}
\newcommand\bB{\overline B}

\title[ %FE I did put back radially dependent
Large Scale Dynamos in Cylinders]{
 Radially Dependent Large Scale   Dynamos in Global
     Cylindrical  Shear Flows and the  Local Cartesian Limit}  
\author[F. Ebrahimi, E. G. Blackman]{F. Ebrahimi$^{1}$, E. G. Blackman$^{2},^{3}$ \\
$^{1}$ Department of Astrophysical Sciences, Princeton University, Princeton, NJ 08543, USA\\
$^{2}$Department of Physics and Astronomy, University of Rochester, Rochester, NY 14627, USA\\
$^{3}$Laboratory for Laser Energetics,University of Rochester, Rochester NY 14623, USA}

\date{\today}
\begin{document}  

\label{firstpage}
\pagerange{\pageref{firstpage}--\pageref{lastpage}}
\maketitle

\begin{abstract}
For  cylindrical differentially rotating plasmas, we study  large-scale magnetic  field generation from finite amplitude non-axisymmetric 
 perturbations  by comparing    numerical simulations
with quasi-linear analytic theory.
When  initiated with a vertical magnetic field of  either zero or finite net flux,  
our global cylindrical  simulations exhibit the magnetorotational instability (MRI) and  large scale dynamo growth of   radially alternating mean fields,  averaged over height and azimuth. This dynamo growth  is explained by our analytic calculations of a  non-axisymmetric fluctuation-induced EMF
that is sustained by azimuthal shear of the fluctuating fields. 
The  standard ``$\Omega$ effect'' (shear of the mean field by differential rotation) is  unimportant.  
For the MRI case, we express the  large-scale dynamo field  as a function of differential rotation.
The resulting radially alternating  large-scale fields may  have implications for angular momentum transport in disks and corona.
%FE I replace it with connect To make contact 
To connect with previous work on large scale dynamos with local linear shear and identify the minimum conditions needed for large scale field growth, we also solve our equations in local Cartesian coordinates.  We find  that large scale  dynamo growth in a linear shear flow  without rotation can be sustained by shear 
plus non-axisymmetric fluctuations--even if not  helical, a seemingly  previously unidentified distinction.
The linear shear flow  dynamo emerges as a more restricted  version of our more general new    global cylindrical calculations. 
\end{abstract}
\maketitle

\section{Introduction}

Astrophysical rotators,  such as stars, galaxies, and accretion disks, commonly show evidence for contemporaneous presence of disordered turbulence  and  magnetic fields ordered on spatial or temporal scales larger than those of the fluctuations. 
Explaining this circumstance has been a long standing challenge.   In situ amplification of  large-scale  magnetic fields via some type of large-scale dynamo 
is  likely but  how these dynamos operate and saturate in each context remains an active subject of research 
(for reviews see \citep{Brandenburg2005,Blackman2015H}).
How such fields grow given the presence  of fluctuations, what are the best analysis methods, and  
what minimum ingredients for growth are needed 
\citep{vishniac97,brandenburg2008,Yousef2008,Heinemann2011,rincon2011,Squire2015b} are topics of active investigation.

Beyond stellar and galactic contexts, evidence for large-scale field growth
is seen in magnetically dominated laboratory plasmas~\citep{ji95,spheromak},
and in local and global simulations 
\citep{Brandenburg1995,ebrahimi2009,Lesur2010,stratified2010,simon2011,Guan2011,Sorathia2012,Suzuki2014}
of the magnetorotational instability (MRI)
\citep{velikhov59,balbus91}.
 Large-scale fields in MRI flows  have been associated with the sustenance of MRI turbulence\citep{lesurdynamo,stratified2010,simon2011} are correlated with the convergence of  Maxwell stress \citep{Guan2011,Nauman2014}, and can influence corona formation \citep{blackman2009}.
For a single MRI mode, large-scale magnetic fields 
generated via an EMF  can  cause  MRI  saturation~\citep{ebrahimi2009}.
In short, the large-scale dynamos of MRI-unstable systems are of interest both as  phenomena on their own, and because 
they may be closely connected to angular momentum transport in accretion disks by  local and nonlocal Maxwell stresses \citep{blackman2015}. In addition to numerical simulations, flow-dominated laboratory experiments are also investigating the MRI MHD unstable systems in Taylor-Couette flow geometry~\citep{goodman02,rudiger03,kageyama04, noguchi02,sisan04,tefani07}.

The  electromotive force  (EMF) from  correlated 
velocity and magnetic field fluctuations is important to all large scale dynamo theories 
 \citep{moffat}. 
In general, correlated fluctuations in the EMF   
facilitate large-scale field amplification and the form that this takes for cylindrical
MHD shear flows  is the focus of the present paper.

As the role of large-scale dynamos and large scale fields for  MRI turbulence and  angular momentum has become increasingly recognized, there is a need for truly  global stratified simulations in the long run to best compare to real astrophysics disks. But  choices must always be made both due to limited computational resources and for  isolating key physical processes that contribute to the global dynamics.    The shearing box model has been the workhorse for simulating MRI turbulence for this purpose-- but it is a local
model and has   limitations associated with boundary conditions, box size and so is a limited model for real  astrophysical disks 
 \citep{regev08,bodo08,blackman2015}. The cylindrical model used here also has some complementary limitations  due to its boundary conditions, but on the other hand provides solutions in a real global domain.  Moreover, as unstratified shearing simulations go, the shearing box in itself might be thought of as a  more restrictive approximation  of a global unstratified cylinder of the sort used here.  
 
 In contrast to  previous studies  of  local Cartesian shearing box addressing large scale field growth from  nonhelical turbulence and linear shear, \citep{brandenburg05_2,Yousef2008,Heinemann2011,Squire2015b},  we  focus on  the more general  large-scale dynamo from the  combination of non-axisymmetric  perturbations and global differentially rotating flows in a cylinder.  We show how the combination of imposed
  non-axisymmetric fluctuations
  and differential rotation, or linear shear of the fluctuating field, is  sufficient to 
source the electromotive force and generate a large scale magnetic field in this cylindrical geometry.   We present the complete quasilinear form of the  EMF and show that it models 
 favorably the results direct numerical simulations (DNS) of the MRI when the magnitude and growth rate of these initial fluctuations in the simulations are used as inputs to the quasi-linear single model analysis. 
 The single mode analysis proves useful in showing explicitly that mode-mode coupling is
not essential for growth, and for identifying which terms in the EMF dominate.  
We also show that these conditions for large scale field growth do not depend on whether the shear profile is favorable or unfavorable to the MRI as long as there is a physically motivated source of
fluctuations.
 
To identify the  minimum requirements for  large scale growth, 
to connect with previous work, and to compare with the global cylindrical  case,
 we  carry out analogous calculations in local Cartesian coordinates.
Comparing  cylindrical and local Cartesian models, we find that 
in each case the fluctuation-induced EMF has 
separate  contributions that depend respectively on 1)  non-uniformity of the radially
sheared non-axisymmetric perturbations 2) the background differential rotation (cylindrical) or linear shear (Cartesian) 3) the mean angular velocity.
 These three vertical EMF terms can separately  generate a large-scale magnetic field.
 We discuss  them in the context of our numerical simulations 
 and the minimum requirements for large scale field growth.
 We find that non-axisymmetric fluctuations plus linear shear 
 OR uniform rotation provide the two most minimal combinations needed for large
 scale field growth in local Cartesian limit.  

%FE added the reference for omega effect
In Sec.~\ref{sec:dns},
we present  evidence of large-scale toroidal fields from global nonlinear MHD DNS 
of the MRI in a cylindrical setup.
We derive the general form of the EMF in the quasilinear approximation in cylindrical geometry in Sec.~\ref{sec:theory1}.
There we also discuss the role of the EMF in field growth (Sec.~\ref{sec:theory1} .2) and consider an example limiting 
 case where radial gradients of the fluctuations are ignored (Sec.~\ref{sec:theory1}.3).
We compare the EMF expression from cylindrical analytics with numerical
calculations in Sec.~\ref{sec:single}. We find that the terms with direct
dependence on mean differential rotation  contribute most to the  dynamo seen in the simulations.
 This is  examined for a specific nonaxisymmetric MRI mode and it is shown that a toroidal large-scale field is directly  generated through the vertical EMF as the result of the coupling of a 
small-scale fluctuations with the differential rotation.  The traditional ``$\Omega$ effect'' (i.e. growth of mean toroidal field from mean poloidal field by shear)\citep{moffat,Parker1979}
is unimportant when the initial mean field is purely vertical.
Visualizations of the field lines from our  nonlinear MRI simulations of the cylinder are presented in Sec.~\ref{sec:vis} to highlight why non-axisymmetric perturbations  are needed for large scale field growth. 
In  Sec.~\ref{sec:theory2} we repeat the quasilinear analysis in local Cartesian coordinates
and derive general forms of the EMF in this geometry.
%We discuss the associated implications for  field growth. By analogy to the 
%examples case discussed in and present the specific simplifying
%example in Sec.~\ref{sec:theory1}.3, we discuss the limiting case in which the $x$ gradients of the fluctuations vanish.
%FE editted the sentence above
We discuss the associated implications for field growth. By analogy to the 
specific simplifying example in Sec.~\ref{sec:theory1}.3, we discuss the limiting case in which the $x$ gradients of the fluctuations vanish (Sec.~\ref{sec:theory2}.2).
Finally, in Sec.~\ref{sec:stable}, we emphasize that large-scale fields can also be generated even if the rotation profiles would imply stability to the MRI,  as long as there is some external supply of non-axisymmetric fluctuations. 
We conclude in Sec.~\ref{sec:sum}, and also present a Table summarizing the ingredients
needed for dynamo action.
%EB4 rewrote the above section

\section{Direct numerical simulations in a cylinder }
%model} 
\label{sec:dns}

We begin with our results from global DNS MHD simulations of the MRI in cylindrical (r, $\phi$, z) geometry using the DEBS~\citep{Debs,ebrahimi2009}  initial-value code to solve the nonlinear, viscous and resistive
 MHD equations
%F.E added code description
\begin{eqnarray}
\frac {\partial \textbf A }{ \partial t } &=& -\textbf{E} = 
S\textbf V\times \textbf B - \eta \textbf J\\
\rho \frac {\partial \textbf V }{ \partial t } &=&
-S \rho \textbf V . \nabla\textbf V + S\textbf J \times
\textbf B +P_m \nabla^2 \textbf V -S \frac{\beta_0}{2}\nabla P  \\
\frac {\partial P }{ \partial t } &=&
-S\nabla \cdot (P \textbf V) - S (\Gamma -1) P \nabla \cdot \textbf V\\
\frac {\partial \rho }{ \partial t } &=&
-S\nabla \cdot (\rho \textbf V) \\
\textbf B &=& \nabla \times \textbf A\\
\textbf J &=& \nabla \times \textbf B
\end{eqnarray}
where the variables, $\rho, P, V, B, J, $, and $\Gamma$ are 
the density, pressure, velocity, magnetic field, current, 
and ratio of the specific heats, 
respectively.
We use the same normalization~\citep{Debs,ebrahimi2009,ebrahimiprl}, where time, radius and velocity are normalized to the outer radius \textit{a}, the resistive diffusion time $\tau_R = a^2/\mu_0\eta$, and the Alfv\'en velocity $V_A = B_0/\sqrt{\mu_0 \rho_0}$, respectively.
The dimensionless parameters, $S= \tau_R V_A/a$ and $P_m$, 
are the Lundquist
number and  the magnetic Prandtl number (the ratio of viscosity to resistivity), respectively. 
the initial state satisfies the equilibrium force balance condition $ \frac{\beta_0}{2}\nabla p  = \rho V_{\phi}^2/r$, where $\beta_0 \equiv 2 \mu_0 P_0/B_0^2$ is  normalized 
to the axis value, and  the initial pressure and density profiles are assumed to be radially uniform and unstratified. Pressure and
density are evolved, however, they remain fairly uniform during the computations.
%F.E 
A no-slip boundary condition is
used for the poloidal flow  and flow fluctuations. The inner and outer radial boundaries are perfectly conducting so that the tangential electric field, the normal component of
the magnetic field, and the normal component of the velocity vanish. The tangential component of 
the velocity is the rotational velocity of the wall. 
The azimuthal ($\phi$) and axial ($z$) boundaries are periodic. 
We assume a radial pressure gradient balances the centrifugal force in equilibrium,
but radial gravity and a radial pressure force  are  interchangeable  for our incompressible, unstratified circumstance.  The pressure gradient,  rather than gravity, is what balances the centrifugal force in cylindrical laboratory experiments designed to test  the MRI  (\cite{goodman02}). 

\begin{figure} 
\includegraphics[width=3.in,height=3.5in]{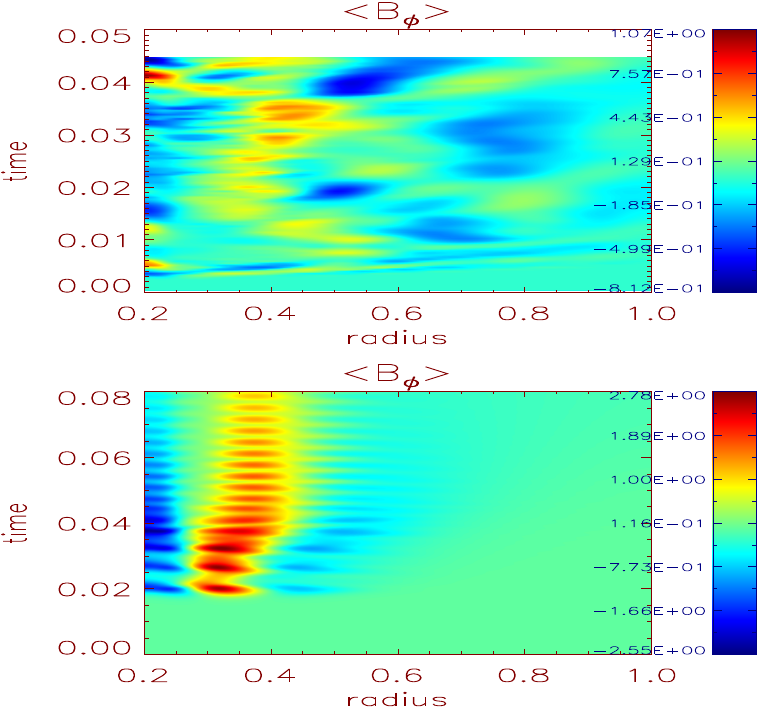}
\caption{The generation of large-scale toroidal magnetic field  (averaged over vertical and toroidal directions) in the zero-net flux simulations in the r-t plane, (a) when all the Fourier modes are included and (b) only one non-axisymmetric mode m=1 is evolved (Rm=3100, Pm=1, $\beta_0=10^5$, $V_0/V_A=31$).}
\label{fig:fig1}
\end{figure}
%F.E added figure
\begin{figure}
\includegraphics[width=3.1in,height=1.8in]{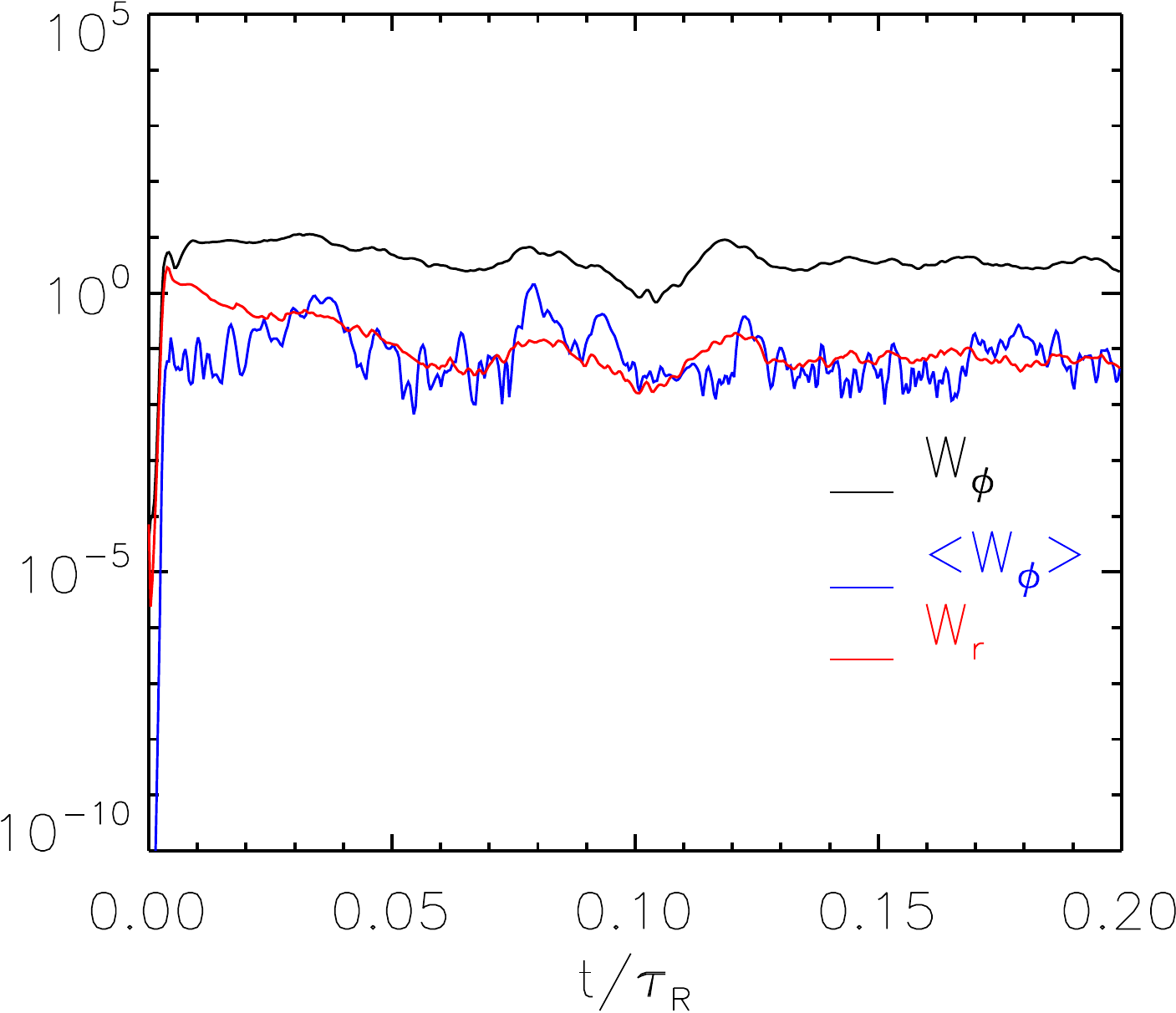}
\caption{ 
Total magnetic energies, $W_{\phi}=1/2 \int{ B_{{\phi}}^2 dr^3}$ and $W_r=1/2 \int{B_{r}^2 dr^3}$, and large-scale toroidal magnetic energy, $ \langle W_{\phi} \rangle=1/2 \int{ \langle B_{{\phi}}\rangle^2 dr^3}$,  vs. time for
net-zero flux 3-D MRI computations.}
\label{fig:fig1_2}
\end{figure}

All variables are decomposed as $f(r,\phi,z,t)=\sum_{(m,k)} \widehat f_{m,k}(r,t) e^{i(-m \phi + kz)}=\lb f(r,t)\rb+\widetilde f(r,\phi\,z,t)$, where $\lb f \rb$ is the mean $(m=k=0)$ component, and $\widetilde f$ is the fluctuating component.  
Mean quantities (indicated  by brackets ($\lb \rb$) or overbars) are  azimuthally and axially averaged, but remain dependent 
on radius (r). 
%F.E added code info
Equations (1-6) are then integrated forward in time using the DEBS code.   
The DEBS code uses a 
finite difference method with a staggered grid for radial 
discretization and pseudospectral method for 
azimuthal and vertical coordinates.  
In this decomposition, each mode satisfies a separate equation of the 
form $\partial \tilde f_{m,k} / \partial t = L_{m,k}  \tilde f_{m,k} +\sum_{(m^{'},k^{'})} N_{m,k,m^{'},k^{'}}$, where $L_{m,k}$ is a 
linear operator that depends on $\tilde f_{0,0}(r,t)$, 
and $N_{m,k,m^{'},k^{'}}$ is a nonlinear term that 
represents the coupling of the mode $(m,k)$ to all other modes $(m^{'},k^{'})$.  (
This latter term is evaluated pseudospectrally.) 
The time advance is a combination of the leapfrog 
and semi-implicit methods~\citep{Debs}.\\
\indent 
 We initiate  simulations with  a   Keplerian flow $\lb V_{\phi}(r)\rb = V_0 r^{-1/2}$ and  uniform  magnetic field $\textbf{B} = B_0 \hat{z}$ (with non-zero initial net-flux) or $\textbf{B} = B_0 \mathrm{sin}(2 \pi (r-r_1)/(r_2-r_1))/r \hat{z}$, 
(with zero-net-flux)
where
$r_1$, $r_2$  are the inner and outer  radii. Fully nonlinear simulations with all  Fourier modes included (with radial, azimuthal and axial
resolutions of $n_r$=220, $0<m<43$ and $-43<n<43$) show that large-scale  magnetic fields are generated [Figure~\ref{fig:fig1}(a)]. In all of our simulations, the initially weak vertical magnetic field, 
gets redistributed, amplified at inner radii and reduced at outer radii. Initially $B_{\phi}$ =0, but a toroidal large-scale (averaged in $\phi$ and $z$)  field grows  via the correlation of non-axisymmetric MRI-induced fluctuations. 
The sustenance of total toroidal and radial magnetic field energies, as well as the large-scale toroidal magnetic energy during the computation for net-zero flux are shown in 
Fig.~\ref{fig:fig1_2}. The radially dependent large-scale field [Figure~\ref{fig:fig1}(a)] is also sustained in time.\\ 
\indent
  The code can also be used to compute the  nonlinear evolution of a single mode evolution
for which  the initial conditions consist of an equilibrium $\langle f(r)\rangle$ plus 
a single mode $\tilde f_{m,k}(r,0) \mathrm e^{( im\phi +ikz)}$ perturbation. The initial amplitude $\tilde f_{m,k}(r,0)$ is a polynomial in $r$ that satisfies the boundary conditions at $r=r_1$ and $r=r_2$.  The initial amplitudes of all other modes are set to zero. 
Only the mode $(m,k)$ is then evolved; however,  the $m=0$, $k=0$ component (the background) is allowed to evolve self-consistently.
The evolution of the background profile $\tilde f_{0,0}$ 
 can affect the evolution of the mode $(m,k)$ and cause the mode to saturate. 
In a fully nonlinear computation, all modes are initialized 
with small random amplitude and are  evolved in time, 
including the full nonlinear term ($N_{m,k,m^{'},k^{'}}$).   
\\
\indent
To facilitate  the analytic investigation of the large-scale field generation, we  carried out 
 nonlinear single-mode (i.e. single value of $m$ and $k$)  simulation of m=1 non-axisymmetric MRI (in which only one fluctuation mode 
and the mean fields self-consistently grow) for zero-net-flux and net-flux configurations. As seen in Fig.~\ref{fig:fig1}(b) at  $t\simeq 0.02$, as the m=1 MRI mode amplitude approaches saturation, a large-scale toroidal field is also generated. \\

\section{Quasilinear analytic calculations of EMFs in a cylinder}
\label{sec:theory1}
\subsection{Deriving the general form of the EMFs}
\indent
To identify the origin of large-scale magnetic field growth in the DNS simulations, we employ quasilinear analytical calculations for the single-mode case. Given  initial fluctuations we calculate the  fluctuation-induced EMF $\emfb\equiv \lb\widetilde {\bf V} \times \widetilde {\bf B}\rb$
from linearized eigenfunctions. The EMF is the source of large scale field growth.
All  averaged correlations are presented in terms of the radial Lagrangian 
displacement  $\mathbf{\xi}_r$ of a fluid plasma element,
 \citep{frieman_rotenberg}, and the mean  quantities.  We assume  perturbed quantities of the form 
$\pmb{\xi}(r, \phi, z, t) = [\xi_r(r),\xi_{\phi}(r),\xi_z(r)]\exp{(\gamma_c t - im\phi+ik_z z)}$ for a cylinder of  outer radius $a$ and 
height  $L = 2 \pi /k$.
In the presence of an equilibrium mean
 flow,  self-adjointness  of the
linear stability problem is lost~\citep{frieman_rotenberg}. Thus, 
for nonaxisymmetric modes (nonzero $m$), the 
eigenvalues $\gamma_c = \gamma +i\omega_r$ and the eigenvectors 
$\xi(r)$ can be complex, where $\gamma$ and $\omega_r$ are the growth rate and the oscillation frequency of the mode, respectively. To  isolate the role of shear flow on the dynamo effect in the quasilinear analytical calculations below,  we impose an initial, uniform $B_0$.

\indent
For a single MRI 
Fourier mode,  the   cylindrical coordinate components of the linearized momentum equation in terms of the Lagrangian 
displacement vector $\mathbf{\xi}$~\citep{Chandrasekhar61} are 
\begin{eqnarray}
 [\bar{\gamma}^2 + \omega_A^2 + 2 r \Omega(r) \Omega^{'}(r)] \xi_r - 2\bar{\gamma} \Omega(r) \xi_{\phi} = - \frac{\partial X}{\partial r}\label{eq:mri1}\\
(\bar{\gamma}^2 + \omega_A^2) \xi_{\phi} + 2\bar{\gamma} \Omega(r) \xi_{r} = \frac{im X }{r}\\
\label{eq:mri2}
(\bar{\gamma}^2 + \omega_A^2) \xi_{z} = -i k_z X,
\label{eq:mri3}
\end{eqnarray}
 where $\bar{\gamma} = \gamma + i \omega_r - i m \Omega(r)$, $X = \widetilde{P} + \widetilde{B}\cdot B_{0} $, 
$\omega_A^2 = k_z^2 B_{0}^2/\rho$ and $\Omega(r) = V_{\phi}(r)/r$ is the angular velocity. 
By including this imposed mean flow in the definition of the Lagrangian displacement vector, the  velocity fluctuations  in an Eulerian frame are given by 
$\widetilde {\bf V} = \partial {\boldsymbol \xi} /\partial t + \nabla \times (\boldsymbol \xi \times \overline{\textbf V}) $, 
with components   $\widetilde V_r = \bar{\gamma} \xi_r $, 
$\widetilde V_{\phi} = \bar{\gamma} \xi_{\phi}- [\frac{\partial {\overline V}_{\phi}}{\partial r}-\frac{{\overline V}_{\phi}}{r}] \xi_r $ and $\widetilde V_z = \bar{\gamma} \xi_z $.

For small resistivity, the  magnetic field
perturbations can be directly related to the displacement via
$\widetilde{\mathbf{B}} = i k_z B_0 {\boldsymbol \xi}$.
Using this along with incompressibility   
and Eqs.(~\ref{eq:mri1}-\ref{eq:mri3}), 
we can eliminate $X$ and 
the azimuthal and vertical displacements can  be written in terms of 
$\xi_r$ as
\begin{equation} 
\xi_{\phi} = \frac{1}{1+m^2/(r^2 k_z^2)}\left[\frac{-2 \Omega(r) \bar{\gamma}}{(\bar{\gamma}^2+\omega_A^2)} \xi_r - \frac{i m}{r^2 k_z^2}(r \xi_r)^{\prime}\right]
\label{eq:xiphi}
\end{equation} 
and
\begin{equation}
\begin{split}
&\xi_{z} = \frac{mk_zr}{(r^2 k_z^2+m^2)}\left(\frac{-2 \Omega(r) \bar{\gamma}}{\bar{\gamma}^2+\omega_A^2} \right)\xi_r \\&+ \frac{i }{r k_z}\left(1- \frac{m^2}{r^2 k_z^2+m^2}\right)(r \xi_r)^{\prime},
\end{split}
\label{eq:xiphi}
\end{equation} 
where the primes indicate radial derivatives.
The quasilinear EMF components  $\emf_z= \lb \widetilde {\bf V} \times \widetilde {\bf B}\rb_{z}={1\over 2} Re(\widetilde V_r^{*} \widetilde{B}_{\phi} - 
\widetilde V_{\phi}^{*} \widetilde{B}_r) $ and $\emf_\phi=\lb\widetilde{\bf  V} \times \widetilde {\bf B}\rb_{\phi} = {1\over 2}Re(\widetilde V_z^{*} \widetilde{B}_r - 
\widetilde V_r^{*} \widetilde{B}_z)$ can now be written in terms of the radial velocity fluctuations $\widetilde V_r = \bar{\gamma} \xi_r $,
\begin{equation}
\begin{split}
&\emf_Z= \lb\widetilde \textbf V \times \widetilde \textbf B\rb_{z} = \emf_{ZVr^{\prime}}+\emf_{Z\Omega^{\prime}}+\emf_{Z\Omega},
\end{split}
\label{eq:vxbz}
\end{equation}
where
\begin{equation}
%\begin{split}
\emf_{ZVr^{\prime}} =   \frac{ m \gamma k_z B_0}{(m^2 + k_z^2 r^2)} \left[\frac{r \widetilde{V_r} \widetilde{V_r^{*}}^{\prime}}{\gamma^2 + \bOmega^2(r)} \right].
%\end{split}
\label{eq:vxbz1}
\end{equation}

\begin{equation}
%\begin{split}
\emf_{Z\Omega^{\prime}} =  - \frac{ m^2 \gamma k_z B_0}{(m^2 + k_z^2 r^2)} \left[\frac{\Omega^{\prime}(r)\bOmega(r) r|\widetilde{V_r}|^2}{(\gamma^2 + \bOmega^2(r))^2} \right], 
%\end{split}
\label{eq:vxbz2}
\end{equation}
\begin{equation}
\begin{split}
&\emf_{Z\Omega}=  \frac{2 \gamma k_z^3 B_0 \Omega(r) r^2  \bOmega}{G^2}(\gamma^2 + \bOmega^2 -\omega_A^2 ) |\widetilde{V_r}|^2 
\\& +  \frac{ m \gamma k_z B_0}{(m^2 + k_z^2 r^2)} \left[\frac{|\widetilde{V_r}|^2}{\gamma^2 + \bOmega^2(r)} \right],
\end{split}
\label{eq:vxbz3}
\end{equation}
and
\begin{equation}
\begin{split}
&\emf_{\phi}=  \lb\widetilde \textbf V \times \widetilde \textbf B\rb_{\phi} = \emf_{\phi Vr^{\prime}}+\emf_{\phi \Omega^{\prime}}+\emf_{\phi \Omega},
\end{split}
\label{eq:vxbt}
\end{equation}
where
\begin{equation}
%\begin{split}
\emf_{\phi Vr^{\prime}} =  \left[\frac{\gamma B_0}{r}-\frac{ m^2 \gamma B_0}{r(m^2 + k_z^2 r^2)}\right] \left[\frac{(r \widetilde{V_r} \widetilde{V_r^{*}}^{\prime})}{\gamma^2 + \bOmega^2(r)} \right],
%\end{split}
\label{eq:vxbt1}
\end{equation}
\begin{equation}
%\begin{split}
\emf_{\phi \Omega^{\prime}} = - \left[\frac{\gamma B_0}{r}-\frac{ m^2 \gamma B_0}{r(m^2 + k_z^2 r^2)}\right] \left[\frac{\Omega^{\prime}(r)\bOmega(r) r|\widetilde{V_r}|^2}{(\gamma^2 + \bOmega^2(r))^2} \right], 
%\end{split}
\label{eq:vxbt2}
\end{equation}

\begin{equation}
\begin{split}
&\emf_{\phi \Omega}= \frac{2 \gamma k_z^2r B_0  \bOmega m \Omega(r)}{G^2}(\omega_A^2 - \gamma^2-\bOmega^2) |\widetilde{V_r}|^2  
\\&+ \left[\frac{\gamma B_0}{r}-\frac{ m^2 \gamma B_0}{r(m^2 + k_z^2 r^2)}\right]\left[\frac{|\widetilde{V_r}|^2}{\gamma^2 + \bOmega^2(r)} \right], 
\end{split}
\label{eq:vxbt3}
\end{equation}
where
\begin{equation}
G^2 = (m^2+k_z^2 r^2)[\gamma^2 + \bOmega^2(r)][4 \gamma^2  \bOmega^2(r) + (\gamma^2 + \bOmega^2(r) +\omega_A^2)^2]
\end{equation} 
and $ \bOmega(r) = m \Omega(r) - \omega_r$. 
In an ideal  MHD cylindrical plasma, Eq. ~(\ref{eq:vxbz}) provides the complete quasilinear form of the vertical fluctuation-induced  EMF  in terms of radial perturbations. The first term on the RHS, $\emf_{ZVr^{\prime}}$, depends on the non-uniformity of the radial displacement of the mode. The second term $\emf_{Z \Omega^{\prime}}$, which depends on the differential rotation $\Omega^{\prime}(r)$ is sufficient to directly produce a nonzero fluctuation-induced dynamo term. 
The free energy source  
$\frac{d\Omega^2}{dlnr}$  appears in this term.
  The third term, $\emf_{Z \Omega}$, 
  shows the dependence of the vertical EMF on angular velocity. 

  The linearized cylindrical solutions ($\gamma$ and $\xi_r$) for nonaxisymmetric flow-driven and MRI modes have been previously examined \citep{bondeson87,ogilvie96,keppens2002}. Here we do not solve the eigenvalue problem to find the $\xi_r$  for nonaxisymmetric modes but
  (in Sec. \ref{sec:single},)
    extract the linearized solutions directly from DNS for a single mode 
  %during  he linear phase 
  and verify the  quasilinear forms of EMFs.

The above quasilinear EMF terms allow us to 
identify the source of  large scale magnetic field growth in a rotating plasma with or without radially sheared
 %helical
  non-axisymmetric 
  perturbations. 

\subsection{EMFs and Large Scale Field Growth }

The dominance of the fluctuation-induced quasilinear $\emf_Z$  in the generation of the large-scale toroidal magnetic field can be seen by  examining the mean (averaged in $\phi$ and $z$) toroidal component of the induction equation, ignoring resistivity. This equation is
\begin{equation}
%\begin{split}
\frac{\partial \overline{\textbf B_{\phi}}}{\partial t} = - \frac{\partial \emf _z}{\partial r}
% - \frac{\partial \lb(\widetilde \textbf V \times \widetilde \textbf B)\rb _z}{\partial r}
 + (\overline{\textbf B} \cdot \nabla) \overline{\textbf V}|_{\phi} - (\overline{\textbf V} \cdot \nabla) \overline{\textbf B}|_{\phi}.
%\end{split}
\label{eq:induction1}
\end{equation}
Since there is neither a mean radial  magnetic field
${\overline B}_r$,  nor velocity field ${\overline V}_r$, so the second and third terms on the right of  Eq. (\ref{eq:induction1}) vanish. Note that the second term on the right of Eq. (\ref{eq:induction1}) is the traditional "$\Omega$ effect" which thus vanishes for 
our setup and averaging procedure. The shear (differential rotation) does enter  through $\emf_z$,
and the first term on the right of Eq.~(\ref{eq:induction1}) is the dominant term.
 
Keeping only the first term on the right of Eq.~(\ref{eq:induction1}) 
($\frac{\partial \overline{\textbf B_{\phi}}}{\partial t} = - \frac{\partial \emf _z}{\partial r}$) 
we then see that the three terms on the right of Eq. ~(\ref{eq:vxbz}) 
provide distinct paths for large scale fields to grow:
(1) Radially sheared 
%FE replaced the ``first two'' above  with three
non-axisymmetric 
perturbations i. e., the first term in Eq. ~(\ref{eq:vxbz}), $\emf_{ZVr^{\prime}}$, proportional to the non-uniformity of the radial displacement of the nonaxisymmetric perturbation
(2) Uniform non-axisymmetric  perturbations (stable or unstable) but with background shear in the angular flow i. e. the last two terms in Eq. ~(\ref{eq:vxbz}).
We emphasize that all three terms on the right of  $\emf _Z$  (Eq. ~\ref{eq:vxbz}) vanish explicitly  for axisymmetric modes (m=0 modes, $\bOmega=0$), as  axisymmetric modes are purely growing or decaying (i.e. $\omega_r=0$) \citep{Chandrasekhar61}. 

\indent

 The  induction equation for the vertical mean field is
 \begin{equation}
%\begin{split}
\frac{\partial \overline{\textbf B_{z}}}{\partial t} =  \frac{\partial \emf_{\phi}}{\partial r} + (\overline{\textbf B} \cdot \nabla) \overline{\textbf V}|_{z} - (\overline{\textbf V} \cdot \nabla) \overline{\textbf B}|_{z}.
%\end{split}
\label{eq:induction2}
\end{equation}
Here  again, as in equation (\ref{eq:induction1}),  the last two terms on the right vanish and  
the mean field  evolves only through the EMF term   $\partial \emf_{\phi}/dr$.
For this equation,  axisymmetric modes can contribute through the radial variations of $\widetilde V_r$ in $\emf_{\phi Vr^{\prime}}$ (Eq.~\ref{eq:vxbt}) and  the last term in Eq.~(\ref{eq:vxbt3}) to give 
\begin{equation}
 \emf_{\phi (m=0)}=  \left(\frac{B_0}{\gamma}\right) [\widetilde{V_r} \widetilde{V_r^{*}}^{\prime} + |\widetilde{V_r}|^2/r],\\
 \emf_{Z (m=0)} = 0
\label{eq:m0}
\end{equation}
but only non-axisymmetric modes allow evolution of
{\it both} ${\overline B}_z$ and  ${\overline B}_\phi$ through the EMF terms.

\begin{figure}
\includegraphics[width=3.in,height=3.in]{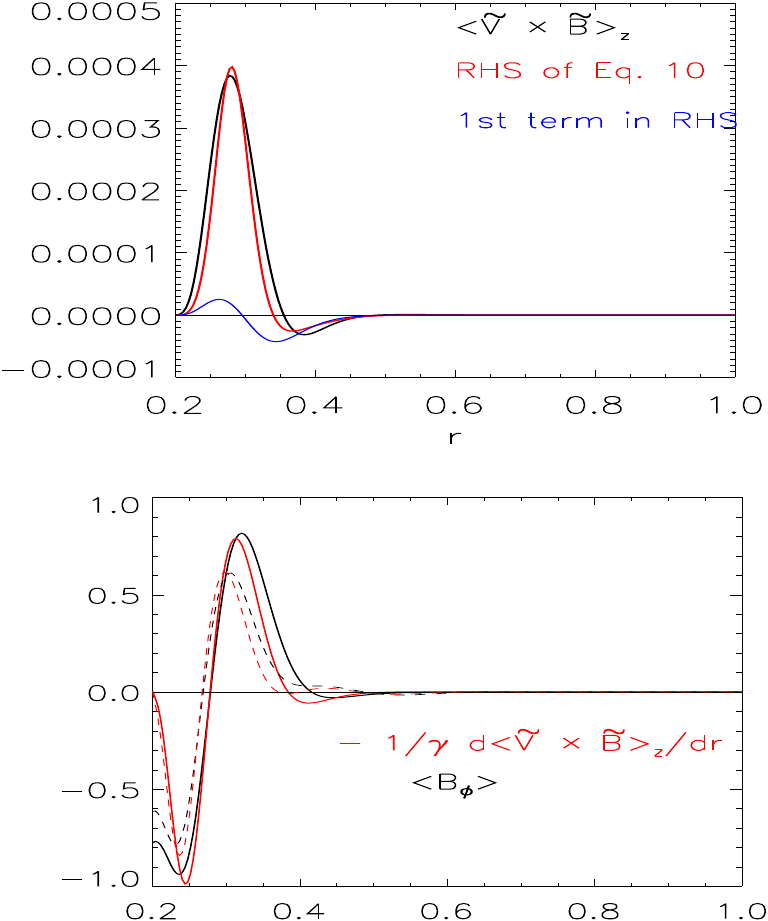}
%\vspace{-45mm}
%FE2 added the info about fluctuations in the caption
\caption{The profiles of (a) (in black) direct numerical calculations of total vertical EMF term, LHS of Eq.~(\ref{eq:vxbz});  (in red)  RHS side of Eq.~(\ref{eq:vxbz}) based on the quasilinear calculations; (in blue) the first term in Eq. (\ref{eq:vxbz}). Linear growth rate of the mode at the early phase $\gamma \tau_{orbit} \sim 0.9$ and $\omega_r \tau_{orbit} \sim 2.2$ have been used for the calculations of RHS, (b) $\lb B_{\phi}\rb$  and  $-{1\over \gamma} \frac{\partial \emf_z}{\partial r}$ at $t/\tau_{orbit} =16$ and $ t/\tau_{orbit} =10$ during the growth of m=1 MRI mode from DNS with non-zero net flux (solid lines) and zero net flux (dashed lines), respectively. The dimensionless  magnetic and velocity fluctuations at this time of DNS are  $|\widetilde{B_{\phi}}/B_0| \sim 0.06$ and $|\widetilde{B_{r}}/B_0|= {k_z a |\widetilde{V_{r}}/V_A| \over k_0 \gamma \tau_A }\sim 0.015$, and
using DNS values of $\gamma \tau_A \sim 12$, $|\widetilde{V_{r}}/V_A| \sim 0.015$, ($V_0/V_A =16$). }
\label{fig:fig2}

\end{figure}

\subsection{EMFs appropriate for MRI driven fluctuations when $k_r=0$}

To simplify pinpointing the dominant contributions to the EMF from the quasi-linear theory appropriate for MRI driven fluctuations for an initially  vertical field, 
%FE added below ``from DNS in the following section''
we  assume $\widetilde{V_r^{\prime}}=0$ (in the limit of $k_r=0$). This is justified since  Fig.~\ref{fig:fig2}(a) from DNS in the following section shows that the term arising from this radial gradient is subdominant.  We can now solve the quasilinear equations for $\emf_z$ without knowing the exact form of the global eigenfunctions. 
For large growth rates  (i.e. $\gamma^2 , {\overline\Omega}^2 > \omega_A^2$), Eq.~(\ref{eq:vxbz}) then reduces to
\begin{equation}
\begin{split}
&\emf_Z (r) \sim   \frac{  \gamma k B_0}{(m^2 + k^2 r^2)} 
\left[m - \frac{m^2 \Omega^{\prime}(r)\bOmega(r) r }{\gamma^2 +\overline \Omega^2}+ \frac{2k^2 {V}_{\phi} r \overline \Omega}{\gamma^2 +\overline \Omega^2}\right]\\&\times (\frac{|\widetilde{V_r}|^2}{\gamma^2 + \bOmega^2(r)}).
\label{eq:vxbz2}
\end{split}
\end{equation}
This vertical EMF  for a single nonaxisymmetic mode  and the mean flow (e.g. Keplerian) with $\gamma \sim \Omega_0$ (where $\Omega_0$ is the angular frequency at the inner radial boundary)
 in the  $k_zr > m$ and $\gamma^2 > {\overline\Omega}^2$ limits are then related by
$\emf_z\sim - \frac{m^2 B_0}{k_z r^2 \gamma^3} \frac{d\Omega(r)^2}{dlnr}|\widetilde{V_r}|^2 +  2 m kB_0 \frac{\Omega(r)\bOmega(r)}{\gamma^3} |\widetilde{V_r}|^2\equiv Q(r) \widetilde{V_r}|^2.$
Using this equation and $\overline B_{\phi} \sim- {1\over \gamma} \frac{\partial \emf_z}{\partial r}$ 
 in the limit of constant $\widetilde V_r$ and constant mean vertical magnetic field $B_0$,
the large-scale toroidal magnetic field can be written as
 \begin{equation}
 {\overline B}_{\phi} (r) \sim - \frac{m^2 B_0}{k_z \gamma^4} \left[\frac{1}{r^2}\frac{d\Omega(r)^2}{dlnr}\right]^{\prime}|\widetilde{V_r}|^2 + \frac{2mkB_0}{\gamma^4}  \left[\Omega(r)\bOmega(r)\right]^{\prime}|\widetilde{V_r}|^2,
\label{eq:btor}
 \end{equation}
showing a direct relationship between differential rotation and the generation of large-scale magnetic field.\\

\section{Comparing  theory with DNS  of a nonaxisymmetric mode}
\label{sec:single}

From DNS of a nonaxisymmetric mode $m=1$, $k_z a$=12 in cylindrical model (Sec.~\ref{sec:dns}),  
we  evaluate 
 $ \lb\widetilde \textbf V \times \widetilde \textbf B\rb_{z} $  and compare  it to quasilinear calculations of the right  side (RHS) of Eq.~(\ref{eq:vxbz}) in terms of radial velocity fluctuations $ \widetilde V_r =\bar{\gamma}\xi_r $. For 
the RHS of Eq.~(\ref{eq:vxbz}),  the radial velocity fluctuations and the eigenvalues from DNS are inserted into the analytical forms.  Fig.~\ref{fig:fig2}(a) 
 shows  good agreement between these two calculations.
Fig.~\ref{fig:fig2}(a) also shows that the first term on the RHS of  Eq.~(\ref{eq:vxbz}) is subdominant to the last two terms
 which depend  on the mean flow. 

\begin{figure}
\includegraphics[width=2.8in,height=3.3in]{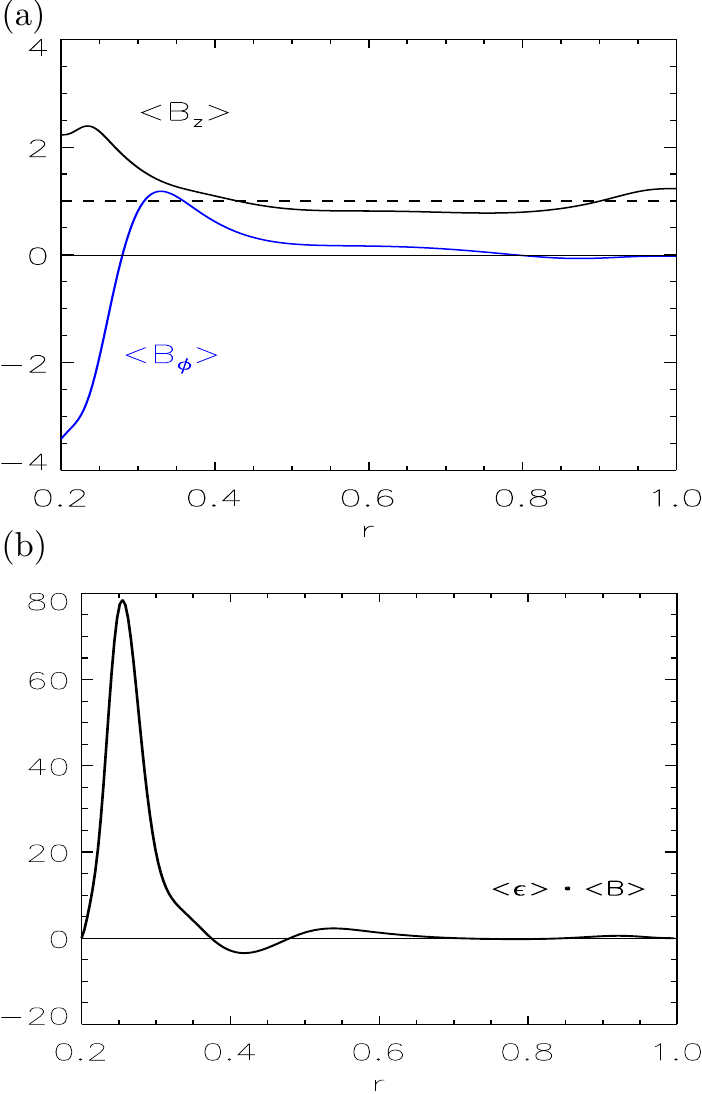}
\caption{Radial profiles of time-averaged (b) saturated large-scale fields $\lb B_z \rb, \lb B_{\phi}\rb$, (b) $\emf \cdot \bbB$ ( = S $\lb\widetilde \textbf V \times \widetilde \textbf B\rb$ $\cdot \lb \textbf{B} \rb$ from DNS of Eq. (1-6) during nonlinear evolution of m=1 MRI.}
\label{fig:fig3_2} 
\end{figure}

We have verified the dominance of the first term on the right of Eq. (\ref{eq:induction1})
 from  DNS of a single-mode m=1 MRI. The large-scale ${\overline B}_{\phi}$  starts to grow, even when initially zero, as the instability develops. Figure~\ref{fig:fig2}(b) shows  $\lb B_{\phi} \rb $  as computed from the DNS during the linear phase of single-mode simulations  with non-zero net flux 
 and the first term on the RHS of  Eq.~(\ref{eq:induction1}),
   right before the saturation, as also measured from the DNS.
As seen, the mean toroidal field is correlated with, and directly generated by the vertical EMF.  
Similarly, the mean $B_\phi$ generated in the net-zero flux simulations shown in Fig.~\ref{fig:fig1} is also correlated with the vertical EMF.  Fig~\ref{fig:fig2}(a) shows that the main contribution to the EMF comes  from the last two terms of  Eq.~(\ref{eq:vxbz}).  
Thus $B_{\phi} \sim- {1\over \gamma} \frac{\partial \lb (\widetilde \textbf V \times \widetilde \textbf B)_{z}\rb}{\partial r}$
 is directly dependent on the shear-flow ($V_{\phi}^{\prime}$) or differential rotation ($d\Omega/dr$) in the presence of a finite amplitude fluctuation. 

In addition to the  EMF components themselves, 
the magnetic field-aligned EMF  plays important role in the quasi-linear regime as the mode starts to saturate.    Evolution of both components (toroidal and vertical) of the large-scale magnetic field is only possible for  nonaxisymmetric fluctuations  because only in this case are both $\emf_z$ and $\emf_{\phi}$  (Eq.~\ref{eq:vxbz},~\ref{eq:vxbt}) non-zero. 
The generation of the large-scale toroidal field is  due to the vertical EMF and the redistribution (and amplification around r=0.2-0.4) of ${\overline B}_z$ is due to the nonzero toroidal EMF.
Figure~\ref{fig:fig3_2} shows the profiles of time-averaged saturated large-scale
toroidal and vertical fields as well as the  EMF parallel to $\overline{\bf B}$. 

Simulations with  axisymmetric fluctuations also show the amplification of vertical large-scale field ($B_z$) but without a generation of $B_{\phi}$. The amplification of $B_z$ from axisymmetric modes (Eq.~\ref{eq:m0}), which may contribute to the mode saturation in a cylinder~\citep{ebrahimi2009} results from the curvature terms and is absent for channel modes in a local Cartesian model, 
 as  discussed below. 

\begin{figure}
 \includegraphics[height=5.in]{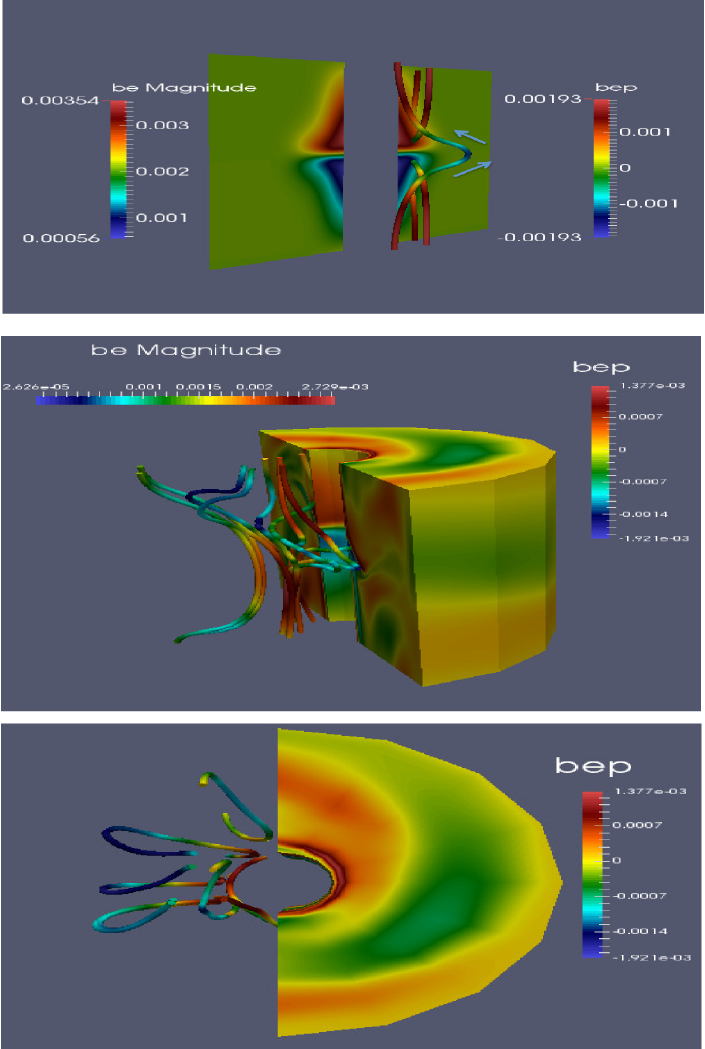}
\caption{Field line visualizations
 during nonlinear MRI simulations in a cylinder (a) 2-D (b) 3-D when m=1 mode perturbation is dominant (c) toroidal top view in 3-D shows the twisting of the field lines.  
}
\vspace{-6mm}
\label{fig:fig3}
\end{figure}

\section{Visualizations from Numerical Simulations}
\label{sec:vis}

The physical picture of generating $\langle B_\phi \rangle$ can further be examined through
comparing  visualizations of the field lines without and with nonaxisymmetric MRI perturbations in 2-D and 3-D as shown in Fig.~\ref{fig:fig3}. 
 For  toroidal m=0 perturbations, weak vertical magnetic 
field lines are toroidally stretched (Fig.~\ref{fig:fig3}(a)) according to the second term in Eq.~(\ref{eq:induction1}). However due to toroidal symmetry, 
$\langle B_\phi \rangle=0$ as the positive and negative contributions from the perturbations  remain on the same vertical surface, 
and only the mean vertical field is amplified through $\partial \emf_{\phi}/dr$.~\citep{ebrahimi2009}. In the presence of  nonaxisymmetric perturbations in 3-D nonlinear simulations however, the field lines are stretched and \textit{twisted}
 (Fig.~\ref{fig:fig3} b,c). As a consequence, $\langle B_\phi \rangle\ne 0$ since now the 
 oppositely signed toroidal field contributions from perturbations are displaced radially from one another.
 Since $\langle B_r\rangle $ is zero, the standard ``$\Omega$ effect'' [$\langle B_r \rangle V_{\phi}^{\prime}$] contribution in Eq.~(\ref{eq:induction1}) is zero.

\section{Local Cartesian Quasilinear analytic calculations of EMFs }
\label{sec:theory2}
\subsection{General form of the EMFs}

Here, we present the analog  to Eqs.~(\ref{eq:mri1}-\ref{eq:vxbz}) in  local Cartesian coordinates $(x,y,z)$ in a frame rotating with fixed angular velocity  $ \mathbf{\Omega}  =\Omega_0 \textbf e_z $
 and a linear shear velocity of $\textbf V_0 = V_y(x) \textbf e_y $. We again assume a vertical field $B_0{\hat \bf z}$ 
 but now assume  perturbed 
 velocity and magnetic field
  of the form  $\pmb{\xi}(x, y, z, t) = [\xi_x(x),\xi_{y}(x),\xi_z(x)]\exp{(\gamma_c t - ik_yy+ik_z z)}$, where, $\gamma_c = \gamma +i\omega_r$.  
  In this rotating  unstratified system, we  include the Coriolis force and 
  the centrifugal force,  and   again assume that in equilibrium, the latter 
  is canceled by the radial pressure gradient.
  (As for the cylindrical case,  the role of gravity vs. radial pressure gradient are 
  interchangeable for our incompressible, unstratified case. ) 
  The   momentum equation $\rho \frac {\partial \textbf V }{ \partial t } =
- \rho \textbf V . \nabla\textbf V + 2 (\mathbf{\Omega}\times \textbf V) + \mathbf{\Omega}\times \mathbf{\Omega} \times{ \bf r}+
 \textbf J \times
\textbf B - \nabla P$
 and the induction equation $\partial_t {\bf B}=\nabla \times ({\bf V}\times {\bf B})$
are linearized in the incompressible limit to give:
\begin{equation}
\begin{split} 
\bar{\gamma}\widetilde{V}_x  - 2 \Omega_0 \widetilde{V}_{y}&=i k_z B_0 \widetilde{B}_x/\rho -\frac{\partial X}{\partial x}  \\
\bar{\gamma}\widetilde{V}_{y} + V_y^{\prime}(x)\widetilde{V}_x + 2 \Omega_0 \widetilde{V}_{x}&=  i k_z B_0 \widetilde{B}_{y}/\rho + ik_y X \\
\bar{\gamma}\widetilde{V}_{z} &=i k_z B_0 \widetilde{B}_{z}/\rho - ik_z X 
\end{split}
\label{eq:momentum}
\end{equation} 
and
\begin{equation}
\begin{split} 
\bar{\gamma} \widetilde{B}_x&= i k_z B_0 \widetilde{V}_x\\
\bar{\gamma} \widetilde{B}_{y}&= V_y(x)^{\prime}\widetilde{B}_x + i k_z B_0 \widetilde{V}_y\\
 \bar{\gamma} \widetilde{B}_{z}&= i k_z B_0 \widetilde{V}_z,
\end{split}
\label{eq:induction}
\end{equation}
where  primes denote variation in x direction, ($e.g. \ \partial/\partial x$) 
Using $\widetilde V_x = \bar{\gamma} \xi_x $, 
$\widetilde V_{y} = \bar{\gamma} \xi_{y}- V_y^{'}(x)\xi_x$ and $\widetilde V_z = \bar{\gamma} \xi_z $, 
Eqs.~(\ref{eq:momentum}) and (\ref{eq:induction})  can be written in terms of the displacement vector as
\begin{eqnarray}
 [\bar{\gamma}^2 + \omega_A^2 + 2 r \Omega_0 V_y^{'}(x)] \xi_x - 2\bar{\gamma} \Omega_0 \xi_{y} = - \frac{\partial X}{\partial x}\label{eq:mri4}\\
(\bar{\gamma}^2 + \omega_A^2) \xi_y + 2\bar{\gamma} \Omega_0 \xi_x = i k_y X \\
\label{eq:mri5}
(\bar{\gamma}^2 + \omega_A^2) \xi_{z} = -i k_z X.
\label{eq:mri6}
\end{eqnarray}

Analogous to the cylindrical case (Sec.~\ref{sec:theory1}), from these sets of equations, the quasilinear vertical EMF  in terms of the Eulerian velocity fluctuations ($\widetilde {\bf V} = \partial {\boldsymbol \xi} /\partial t + \overline{\textbf V} \cdot \nabla {\boldsymbol \xi} - { \boldsymbol \xi}\cdot \nabla \overline{\textbf V}$), $\widetilde V_x = \bar{\gamma} \xi_x $ is reduced to:

\begin{equation}
\begin{split}
&\emf_z= \lb\widetilde \textbf V \times \widetilde \textbf B\rb_{z} = \emf_{ZVr^{\prime}}+\emf_{Z (shear)}+\emf_{Z\Omega},
\end{split}
\label{eq:vxbzc}
\end{equation} 
where
\begin{equation}
%\begin{split}
\emf_{ZVx^{\prime}} =   \frac{ k_y\gamma k_z B_0}{(k_y^2 + k_z^2)} \left[\frac{\widetilde{V_x} \widetilde{V_x^{*}}^{\prime}}{\gamma^2 + \vartheta^2(x)}\right],
%\end{split}
\label{eq:vxbzc1}
\end{equation}
\begin{equation}
%\begin{split}
\emf_{Z (shear)} =  - \frac{ k_y\gamma k_z B_0}{(k_y^2 + k_z^2)} \left[\frac{k_y V_y(x)^{\prime}(k_yV_y(x) - \omega_r)}{(\gamma^2 + \vartheta^2(x))^2}\right]|\widetilde{V_x}|^2, 
%\end{split}
\label{eq:vxbzc2}
\end{equation}
and
\begin{equation}
\begin{split}
&\emf_{Z\Omega0}=  
%+
 \frac{2 \gamma k_z^3 B_0 \Omega_0 \vartheta(x)}{G^2}(\gamma^2+{\vartheta(x)}^2 -\omega_A^2 ) |\widetilde{V_x}|^2,
\end{split}
\label{eq:vxbzc3}
\end{equation}
where $\vartheta (x) = k_y V_y(x) -\omega_r$, $\bar{\gamma} = \gamma - i \vartheta (x)$,  and $G^2 = (k_y^2+k_z^2)(\gamma^2 + \vartheta^2(x))[4 \gamma^2  \vartheta^2(x) + (\gamma^2 + \vartheta^2(x) +\omega_A^2)^2]$. 
As seen in these equations, in the local Cartesian 
model, the rotation and shear are independent. 

The first contribution $\emf_{ZVx^{\prime}}$ shows how nonaxisymmetric perturbations with  radial shear, even without any explicit mean shear flow
can source a vertical EMF.
In the absence of rotation, the second contribution, $\emf_{Z (shear)}$, shows a direct dependence of vertical EMF on the linear shear. A mean shear flow, $V_y(x)^{\prime}$ combined with a radially uniform non-axisymmertic ($k_y\neq 0$) 
perturbation is sufficient to produce $\emf_{Z (shear)}$. The last contribution
 $\emf_{Z\Omega}$, which vanishes in the absence of angular velocity, shows that a contribution to the vertical EMF can result from  
a finite angular velocity ($\Omega_0$) for non-axisymmetric perturbations.\\

Similarly the azimuthal EMF is given by\\
\begin{equation}
\begin{split}
&\emf_{y}=  \lb\widetilde \textbf V \times \widetilde \textbf B\rb_{y} = \emf_{y Vr^{\prime}}+\emf_{y (shear)}+\emf_{y \Omega_0},
\end{split}
\label{eq:vxby}
\end{equation}
where
\begin{equation}
%\begin{split}
\emf_{y Vx^{\prime}} =  \gamma B_0 \left[1-\frac{k_y^2}{(k_y^2 + k_z^2)}\right] \left[\frac{(\widetilde{V_x} \widetilde{V_x^{*}}^{\prime})}{(\gamma^2 + \vartheta^2(x))} \right],
%\end{split}
\label{eq:vxby1}
\end{equation}
\begin{equation}
%\begin{split}
\emf_{y (shear)} = -\gamma B_0 \left[1 -\frac{k_y^2}{(k_y^2 + k_z^2)}\right] \left[\frac{k_y V_y(x)^{\prime}(k_y V_y(x) -\omega_r)|\widetilde{V_x}|^2}{(\gamma^2 + \vartheta^2(x))^2} \right], 
%\end{split}
\label{eq:vxby2}
\end{equation}
and
\begin{equation}
\begin{split}
&\emf_{y \Omega0}= \frac{2 \gamma k_z^2 B_0 \Omega_0 k_y(k_y V_y(x) -\omega_r)}{G^2}(\omega_A^2 - \gamma^2-\vartheta^2(x)) |\widetilde{V_x}|^2.
\end{split}
\label{eq:vxby3}
\end{equation}

Spatial derivatives of Eq.~(\ref{eq:vxby}) could, in principle, also generate and redistribute  the vertical field $B_z$ due to nonaxisymmetric modes ($k_y\neq 0$). However,  the fastest growing axisymmetric modes--the channel modes ($k_y=k_x=0$)---do NOT contribute in either the vertical or azimuthal EMF (Eq.~\ref{eq:vxbzc} and Eq.~\ref{eq:vxby}) 
obtained above. In contrast, for  the global cylindrical model, even for radially uniform axisymmetric modes ($k_r=m=0$), the last term in Eq.~(\ref{eq:m0}) DOES contribute in the amplification of vertical field. This distinction highlights at least one circumstance in which 
the absence of curvature in the Cartesian model removes a contribution that could be present
in the global rotator.

\subsection{Exact expression for large scale field with only linear shear when $k_x=0$}
A large-scale magnetic field can be generated by the EMF of  Eq.~(\ref{eq:vxbzc}) from any of the independent contributions in ~(\ref{eq:vxbzc1}-~\ref{eq:vxbzc3}).  In the absence of rotation, a large-scale magnetic field, $\overline B_{y} \sim- {1\over \gamma} \frac{\partial \emf_z}{\partial x}$, can directly be generated via a linear flow-shear and a radially uniform non-axisymmetric  ($k_y,k_z\neq 0$, $k_x=0$) perturbation, 
\begin{equation}
 {\overline B}_{y}(x) =  \frac{ k_y k_z B_0}{(k_y^2 + k_z^2)} \left[\frac{k_y V_y(x)^{\prime}(k_yV_y(x) - \omega_r)}{(\gamma^2 + \vartheta^2(x))^2}\right]^{\prime}|\widetilde{V_x}|^2, 
\label{eq:by}
 \end{equation}
This is an exact analytical equation for a large-scale azimuthal magnetic field generated via a linear mean shear-flow and any perturbations with nonzero $k_y$ and $k_z$. 

The large-scale field given in Eq.~(\ref{eq:by}) is consistent with previous studies  of large scale field  growth from the combination of linear shear with randomly forced turbulence
\citep{vishniac97,Yousef2008,Heinemann2011,mitra2012,sridhar2014}. However, our calculations explicitly reveal
the most minimalist conditions needed  for growth in the absence of rotation: a background linear shear and  an imposed non-axisymmetric perturbation with 
nonzero $k_y,k_z$. Helical velocity perturbations are not required.

Generation of ${\overline B}_{y}$ 
in this case of mean shear can be visualized 
by considering an perturbation in the $x$ direction and then considering why both
$z$ and $y$ variations are needed to produce a net field in vertically averaged planes.
 If there were no $z$ variation in the perturbation then the mean shear would produce no toroidal field even before vertically averaging.  And if there  were a vertical variation but
 no  $y$ variation, then the the mean shear would produce  ${\overline B}_y(x)=0$ 
 from vertical
 averaging. 

\begin{figure}
 \includegraphics[width=3.in,height=1.9in]{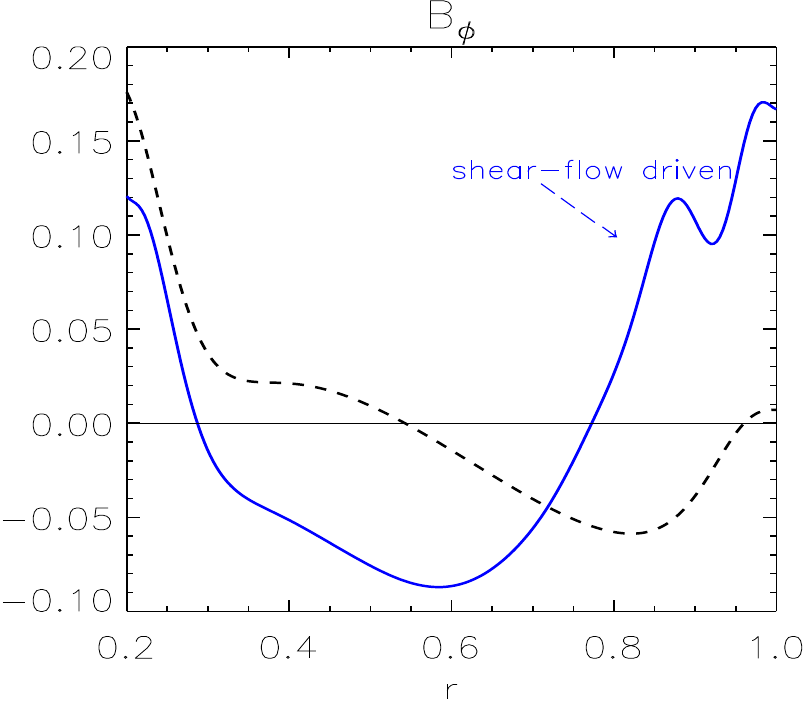}
\caption{$B_{\phi}$ generated with positive shear flow ($V_{\phi}(r)/V_A =80(r/a)^3$) and large initial amplitude forcing (solid line); and only with forcing amplitude varying with radius (dashed line). 
}
\vspace{-6mm}
\label{fig:fig4}
\end{figure}

\section{Large-scale field in the case of stable flow}
\label{sec:stable}

Our quasi-linear theory  imposes fluctuations and background shear as a starting point whereas in DNS,
the fluctuations can directly result from the MRI.  The quasilinear theory shows that the
growth of $\bB_\phi$ in Eq. (\ref{eq:induction1}) via $\partial_r\emf_z$ does not  requires a shear profile favorable to the MRI, just a source of fluctuations and differential rotation of either sign.
To show  this, we numerically  computed  the large-scale field growth
from quasilinear theory  by initializing single mode fluctuations (${\tilde f}_{m,k}(r,0)$ with a polynomial dependence on $r$) in the simulations and forcing amplitudes of $10\%$ on top of a 
 stable equilibrium flow $d\Omega/dr >0$.
 
As the fluctuation energy decays, $\bB_\phi$ grows  via $\emf_z$ from (Eq.~\ref{eq:vxbz}). Figure~\ref{fig:fig4} shows the large-scale toroidal field generated 
 using $\bV_{\phi}(r)/\bV_A =80(r/a)^3$. The profiles are time-averaged during the decay phase.
 In Figure~\ref{fig:fig4} we  have also shown the  case when  $\bB_{\phi}$ is generated by forcing only with the same radially dependent 
fluctuations but in the absence of  mean shear.
 For this latter case, $\emf_z$ in Eq.~(\ref{eq:vxbz}) is then dominated by the first term on the right.
Comparing the two cases, we see that for small $r/a$ the case with only radius dependent fluctuations  (dashed line in Fig.~\ref{fig:fig4})  also captures the  growth of $B_{\phi}$ 
as for the case with both fluctuations and shear.  But for radii of  large shear  
$r/a>0.75$, the flow-dependent EMF terms (the last two terms of Eq.~\ref{eq:vxbz}) dominate.

\begin{table*}
\centering
\begin{tabular}{| c | c |c | c |c | c | c|c|}
\hline \hline
  {\bf EMF radial derivative}  &\multicolumn{3}{|c|}{\textbf{Restriction Maintains Finiteness?}}\\
\hline \hline
 global cylinder (with $\Omega >0; \Omega' \ne 0; k \ne 0; k_r\ne 0;  m>0$) &$k_r= 0 $    &  $ \Omega^{\prime}(r) = 0 $ &  $ m =0$\\\hline \hline
 $ \emf_{Z}'(r) \ne 0$& yes & yes & no  \\
  $\emf_{\phi}'(r)\ne 0$ &yes & yes& yes \\
  $ \emf_{\phi}'(r) $ and $\emf_{Z}'(r) \ne 0$ (dynamo)& yes &yes &no \\
 \hline\hline
local Cartesian 
(with $\Omega_0 >0; V_y' \ne 0; k\ne 0; k_x\ne 0;  k_y>0$)
 & $ k_x= 0$  & $V_y^{\prime}(x) = 0$ &$k_y = 0$  \\ \hline \hline
 $ \emf_{Z}'(x)\ne 0$ & yes& no & no \\
 $  \emf_{y}'(x)\ne 0$ &yes & no & no\\
$ \emf_{y}'(x) $ and $\emf_{Z}'(x) \ne 0$ (dynamo)& yes &no &no \\
% $$ &yes & no& no  & yes & yes  \\
\hline \hline
local Cartesian (with $\Omega_0 =0; V_y' \ne 0;k\ne0;  k_x\ne 0; k_y>0$) &$ k_x= 0$  & $V_y^{\prime}(x) = 0$ &$k_y = 0$  \\ \hline \hline
 $ \emf_{Z}'(x)\ne 0$ & yes& no & no \\
 $  \emf_{y}'(x)\ne 0$ &yes &  no & no\\
$ \emf_{y}'(x) $ and $\emf_{Z}'(x) \ne 0$ (dynamo)& yes  &no &no \\
% $$ &yes & no& no  & yes & yes  \\
\hline \hline
\end{tabular}
\caption{Table summarizing the  minimal ingredients 
needed to maintain the finiteness of the two  components of the EMFs separately
and together.  The first column indicates
three general cases: the global cylinder and two Cartesian cases. 
To read the table for each of these cases, consider  the global cylinder case as an example:
the second row of the  first column indicates general ingredients that our global cylinder could have
(rotation $\Omega >0$, differential rotation  $\Omega' \ne 0$, and general
perturbations with $k \ne 0, k_r \ne 0$, and   $m>0$ for non-axisymmetry). 
 Each of the subsequent columns indicate  a  restriction that reduces this generality. The entries   "yes" or "no" within these columns indicate whether the 
 corresponding quantity in the first column is finite when that restriction  is imposed.  
 For each of the three  cases (global cylinder,  Cartesian with $\Omega_0>0$ and Cartesian with $\Omega_0=0$) listed,
 a first and second row provide the information on the finiteness of each component of the 
 and  the third row provides the information on the finiteness of both components together.
 This latter circumstances is needed to supply the large scale dynamo.  Note that if either of the two separate components has a "no" entry then that also implies that both together cannot be finite.
 The conditions for  finiteness (and thus  dynamo action) are different in the two geometries
 because curvature terms are absent in the Cartesian approximation.}
\end{table*}

\section{Summary and conclusions}
\label{sec:sum}

 In summary,  we have shown from both numerical simulations and 
 semi-analytic quasi-linear theory   how radially alternating large-scale toroidal fields averaged vertically and azimuthally can be generated from
MHD flow-driven fluctuations. These fields are  found in  MHD DNS for  both zero-net-flux and non-zero-net-flux initial configurations in both the quasi-linear regime and the fully saturated non-linear regime.

Given nonaxisymmetric fluctuations (with nonzero vertical and azimuthal perturbations), 
we calculated the contributions to the  quasilinear fluctuation-induced EMFs in both cylindrical and Cartesian coordinates.  
We have separated the derivation of the global and local models 
 so that the reader can study them separately. 

  We have not presented physical interpretations of all circumstances that can lead to growth from these equations, but have provided the general forms of the EMFs and identified the minimum
  requirements for growth.    Table 1 summarizes  these requirements for a nonzero EMF and large-scale field growth in both cylindrical and Cartesian models.
    In general, we find a direct relationship between dynamo generating EMFs  and differential rotation in the cylindrical model, or linear shear in the local Cartesian model.  
  The vertical EMF  associated with fluctuations in the presence of an initial vertical field is sufficient to generate an azimuthal
    large-scale field  for non-zero differential rotation (in rotating system) but requires
    non-zero flow shear in the local Cartesian model for  a non-rotating system.  
    
    Table 1 also highlights that due to the absence of curvature terms, the local Cartesian model
   is more restrictive for field  growth than global cylindrical model.
    According to our Cartesian EMF calculations, the fastest growing channel modes (with $k_y=k_x=0$)~\citep{goodmanxu} found in shearing box simulations
    do NOT contribute to  the EMFs in the local approximation (and thus the saturation of these modes) but the analogous modes
    can amplify large-scale fields and  contribute to MRI saturation in global  cylindrical simulations~\citep{ebrahimi2009}.

    In the case of a large scale flow-driven instability, the free energy source from the large scale motion can be the source of the needed fluctuations.  
 For the global cylinder, we have indeed found  explicit dynamo generation of  $\langle B_\phi \rangle$ 
 from DNS where the MRI  produces a  fluctuation-induced vertical EMF $\emf_z$.
 The DNS provide properties of the  fluctuations  that we use 
 as inputs to a quasi-linear calculation of the dynamo growth for a single mode.
 The DNS  large scale field growth and the associated quasi-linear dynamo calculations 
 are in reasonable agreement.
 Our study  of the single mode evolution its correspondence with DNS 
 highlights that  "turbulence" (defined as non-linear mode
  coupling) is not actually essential for the large-scale field growth and that  insight is gained 
  even  from
  single mode analyses.
 
  Our results also show that  the traditional ``$\Omega$ effect'' of shear on the mean field is absent when the initial mean field is vertical and the averaging is over vertical surfaces.
  Instead the essential shear operates on the fluctuations.
The field growth can be entirely described by working with the
    EMF directly, non-axisymmetric (though not necessarily helical) 
velocity perturbations are essential for large scale growth as evidenced
    from direct visualization of the field lines in DNS and from the quasi-linear theory. 
  We should also note that in much of the 
MRI  dynamo literature, large-scale fields in shearing boxes
are computed
 via planar averages (and averaged over the direction of the nonuniformaty of the mean flow) leaving mean fields as a function of  z direction.
There, because of the averaging and boundary conditions for the shear box
 ``$\Omega$ effect'' can still survive. Here, our averaging is over vertical surfaces, and not along the direction of mean flow variation.
 An important lesson is that the averaging procedure and boundary conditions have important implications for the dominant contributions to the EMF.
%FE added above about averaging 

 By calculating the complete form of
 EMF for both global cylindrical and Cartesian cases, 
we have demonstrated the minimum ingredients for large scale field growth in both of these
two models.
Our results suggest that the
 quasilinear and nonlinear 
fluctuation-induced EMF may provide  fundamental insight into the growth and sustenance of large-scale dynamo in these flow-driven systems. 
The calculations herein provide  a more general approach to identifying the origin and minimal
ingredients needed for large scale dynamo growth in unstratified rotating and differentially rotating
systems or linearly sheared systems.

Although we leave  a detailed analysis making explicit connections to previous approaches
of  incoherent alpha 
\citep{vishniac97,brandenburg05_2,brandenburg2008,mitra2012,sridhar2014} and or shear current effects as an opportunity for further work, 
we  emphasize two points in this context. First  we have intentionally avoided using the $\alpha$
formalism and worked directly with only the EMF.  Second, we find that the absolute minimum
conditions for  radially dependent large scale field growth are non-axisymmetric velocity fluctuations plus linear shear.
The velocity fluctuations do not need to be helical at any time. In this way our global and local calculations provide a more minimalist set of conditions for growth than  the
that of  a fluctuating kinetic helicity \citep{vishniac97}.  We note however, that in the quasi-linear regime, the large scale magnetic field  does (as a function of radius) 
 (see Fig.{\ref{fig:fig3_2}})  develop a field aligned EMF, 
 which is a source term for sum of  the 
 time derivative of large scale magnetic helicity and  divergence of large scale helicity flux,
as previously confirmed for the global cylindrical case~\citep{ebrahimiprl}.
Here we have not studied  the non-linear/saturating effects of the growth of  small scale magnetic helicity helical fluctuations, nor the
EMF and mean magnetic field correspondence during the nonlinear saturation.
 More detailed calculations for the nonlinear phase of DNS  (by P. Bhat et.al in preparation) do show  a direct correlation of large-scale field with the
 EMF terms in the nonlinear regime that we have presently computed  only in the quasli-linear approximation.

Finally, we note that our large scale fields show radial reversals and these would be sites of current sheets.
If we think toward generalizations to stratified rotators that form coronae, 
only magnetic structures of large enough scale survive buoyant rise into coronae where they can dissipate and transport angular momentum non-locally \citep{blackman2009}.  If our present toroidal field structures and reversal scales survive stratified generalizations, they provide a scale for coronal structures and current sheets that link the large scale field directly to structures associated with coronal transport and dissipation.

\section*{Acknowledgments} 
We thank H. Ji for useful discussions, and thank Axel Brandenburg for useful comments. FE acknowledges  grant support from DOE, DE-FG02-12ER55142 and NSF PHY-0821899 CMSO. This work was also 
facilitated by the MPPC. EB acknowledges   support from 
 NSF-AST-1109285, HST-AR-13916.002, a Simons  Fellowship, and the IBM-Einstein Fellowship Fund at the Institute for Advanced Study during part of this work.


\begin{thebibliography}{}
\makeatletter
\relax
\def\mn@urlcharsother{\let\do\@makeother \do\$\do\&\do\#\do\^\do\_\do\%\do\~}
\def\mn@doi{\begingroup\mn@urlcharsother \@ifnextchar [ {\mn@doi@}
  {\mn@doi@[]}}
\def\mn@doi@[#1]#2{\def\@tempa{#1}\ifx\@tempa\@empty \href
  {http://dx.doi.org/#2} {doi:#2}\else \href {http://dx.doi.org/#2} {#1}\fi
  \endgroup}
\def\mn@eprint#1#2{\mn@eprint@#1:#2::\@nil}
\def\mn@eprint@arXiv#1{\href {http://arxiv.org/abs/#1} {{\tt arXiv:#1}}}
\def\mn@eprint@dblp#1{\href {http://dblp.uni-trier.de/rec/bibtex/#1.xml}
  {dblp:#1}}
\def\mn@eprint@#1:#2:#3:#4\@nil{\def\@tempa {#1}\def\@tempb {#2}\def\@tempc
  {#3}\ifx \@tempc \@empty \let \@tempc \@tempb \let \@tempb \@tempa \fi \ifx
  \@tempb \@empty \def\@tempb {arXiv}\fi \@ifundefined
  {mn@eprint@\@tempb}{\@tempb:\@tempc}{\expandafter \expandafter \csname
  mn@eprint@\@tempb\endcsname \expandafter{\@tempc}}}

\bibitem[\protect\citeauthoryear{Balbus \& Hawley}{Balbus \&
  Hawley}{1991}]{balbus91}
Balbus S.~A.,  Hawley J.~F.,  1991, Astrophys. J., 376, 214

\bibitem[\protect\citeauthoryear{{Blackman}}{{Blackman}}{2015}]{Blackman2015H}
{Blackman} E.~G.,  2015, \mn@doi [Space. Sci. Rev.]
  {10.1007/s11214-014-0038-6}, \href
  {http://adsabs.harvard.edu/abs/2015SSRv..188...59B} {188, 59}

\bibitem[\protect\citeauthoryear{Blackman \& Nauman}{Blackman \&
  Nauman}{2015}]{blackman2015}
Blackman E.~G.,  Nauman F.,  2015, \mn@doi [Journal of Plasma Physics]
  {10.1017/S0022377815000999}, \href
  {http://journals.cambridge.org/article_S0022377815000999} {81}

\bibitem[\protect\citeauthoryear{{Blackman} \& {Pessah}}{{Blackman} \&
  {Pessah}}{2009}]{blackman2009}
{Blackman} E.~G.,  {Pessah} M.~E.,  2009, \mn@doi [\apj]
  {10.1088/0004-637X/704/2/L113}, \href
  {http://adsabs.harvard.edu/abs/2009ApJ...704L.113B} {704, L113}

\bibitem[\protect\citeauthoryear{{Bodo}, {Mignone}, {Cattaneo}, {Rossi}  \&
  {Ferrari}}{{Bodo} et~al.}{2008}]{bodo08}
{Bodo} G.,  {Mignone} A.,  {Cattaneo} F.,  {Rossi} P.,   {Ferrari} A.,  2008,
  \mn@doi [Astronomy \& Astrophysics] {10.1051/0004-6361:200809730}, \href
  {http://adsabs.harvard.edu/abs/2008A%26A...487....1B} {487, 1}

\bibitem[\protect\citeauthoryear{{Bondeson}, {Iacono}  \&
  {Bhattacharjee}}{{Bondeson} et~al.}{1987}]{bondeson87}
{Bondeson} A.,  {Iacono} R.,   {Bhattacharjee} A.,  1987, \mn@doi [Physics of
  Fluids] {10.1063/1.866151}, \href
  {http://adsabs.harvard.edu/abs/1987PhFl...30.2167B} {30, 2167}

\bibitem[\protect\citeauthoryear{{Brandenburg}}{{Brandenburg}}{2005}]{brandenburg05_2}
{Brandenburg} A.,  2005, \mn@doi [\apj] {10.1086/429584}, \href
  {http://adsabs.harvard.edu/abs/2005ApJ...625..539B} {625, 539}

\bibitem[\protect\citeauthoryear{{Brandenburg} \& {Subramanian}}{{Brandenburg}
  \& {Subramanian}}{2005}]{Brandenburg2005}
{Brandenburg} A.,  {Subramanian} K.,  2005, \mn@doi [Phys. Reports]
  {10.1016/j.physrep.2005.06.005}, \href
  {http://adsabs.harvard.edu/abs/2005PhR...417....1B} {417, 1}

\bibitem[\protect\citeauthoryear{{Brandenburg}, {Nordlund}, {Stein}  \&
  {Torkelsson}}{{Brandenburg} et~al.}{1995}]{Brandenburg1995}
{Brandenburg} A.,  {Nordlund} A.,  {Stein} R.~F.,   {Torkelsson} U.,  1995,
  \mn@doi [\apj] {10.1086/175831}, \href
  {http://adsabs.harvard.edu/abs/1995ApJ...446..741B} {446, 741}

\bibitem[\protect\citeauthoryear{{Brandenburg}, {R{\"a}dler}, {Rheinhardt}  \&
  {K{\"a}pyl{\"a}}}{{Brandenburg} et~al.}{2008}]{brandenburg2008}
{Brandenburg} A.,  {R{\"a}dler} K.-H.,  {Rheinhardt} M.,   {K{\"a}pyl{\"a}}
  P.~J.,  2008, \mn@doi [\apj] {10.1086/527373}, \href
  {http://adsabs.harvard.edu/abs/2008ApJ...676..740B} {676, 740}

\bibitem[\protect\citeauthoryear{Chandrasekar}{Chandrasekar}{1961}]{Chandrasekhar61}
Chandrasekar S.,  1961, Hydrodynamic and hydromagnetic stability.
Dover

\bibitem[\protect\citeauthoryear{{Cothran}, {Brown}, {Gray}, {Schaffer}  \&
  {Marklin}}{{Cothran} et~al.}{2009}]{spheromak}
{Cothran} C.~D.,  {Brown} M.~R.,  {Gray} T.,  {Schaffer} M.~J.,   {Marklin} G.,
   2009, \mn@doi [Physical Review Letters] {10.1103/PhysRevLett.103.215002},
  \href {http://adsabs.harvard.edu/abs/2009PhRvL.103u5002C} {103, 215002}

\bibitem[\protect\citeauthoryear{{Davis}, {Stone}  \& {Pessah}}{{Davis}
  et~al.}{2010}]{stratified2010}
{Davis} S.~W.,  {Stone} J.~M.,   {Pessah} M.~E.,  2010, \mn@doi [Astrophys. J.]
  {10.1088/0004-637X/713/1/52}, \href
  {http://adsabs.harvard.edu/abs/2010ApJ...713...52D} {713, 52}

\bibitem[\protect\citeauthoryear{Ebrahimi \& Bhattacharjee}{Ebrahimi \&
  Bhattacharjee}{2014}]{ebrahimiprl}
Ebrahimi F.,  Bhattacharjee A.,  2014, \mn@doi [Phys. Rev. Lett.]
  {10.1103/PhysRevLett.112.125003}, 112, 125003

\bibitem[\protect\citeauthoryear{Ebrahimi, Prager  \& Schnack}{Ebrahimi
  et~al.}{2009}]{ebrahimi2009}
Ebrahimi F.,  Prager S.~C.,   Schnack D.~D.,  2009, Astrophys. J., 698, 233

\bibitem[\protect\citeauthoryear{{Frieman} \& {Rotenberg}}{{Frieman} \&
  {Rotenberg}}{1960}]{frieman_rotenberg}
{Frieman} E.,  {Rotenberg} M.,  1960, \mn@doi [Reviews of Modern Physics]
  {10.1103/RevModPhys.32.898}, \href
  {http://adsabs.harvard.edu/abs/1960RvMP...32..898F} {32, 898}

\bibitem[\protect\citeauthoryear{Goodman \& Ji}{Goodman \&
  Ji}{2002}]{goodman02}
Goodman J.,  Ji H.,  2002, J. Fluid Mech., 462, 365

\bibitem[\protect\citeauthoryear{{Goodman} \& {Xu}}{{Goodman} \&
  {Xu}}{1994}]{goodmanxu}
{Goodman} J.,  {Xu} G.,  1994, \mn@doi [Astrophys. J.] {10.1086/174562}, \href
  {http://adsabs.harvard.edu/abs/1994ApJ...432..213G} {432, 213}

\bibitem[\protect\citeauthoryear{{Guan} \& {Gammie}}{{Guan} \&
  {Gammie}}{2011}]{Guan2011}
{Guan} X.,  {Gammie} C.~F.,  2011, \mn@doi [\apj]
  {10.1088/0004-637X/728/2/130}, \href
  {http://adsabs.harvard.edu/abs/2011ApJ...728..130G} {728, 130}

\bibitem[\protect\citeauthoryear{{Heinemann}, {McWilliams}  \&
  {Schekochihin}}{{Heinemann} et~al.}{2011}]{Heinemann2011}
{Heinemann} T.,  {McWilliams} J.~C.,   {Schekochihin} A.~A.,  2011, \mn@doi
  [Phys. Rev. Lett.] {10.1103/PhysRevLett.107.255004}, \href
  {http://adsabs.harvard.edu/abs/2011PhRvL.107y5004H} {107, 255004}

\bibitem[\protect\citeauthoryear{{Herault}, {Rincon}, {Cossu}, {Lesur},
  {Ogilvie}  \& {Longaretti}}{{Herault} et~al.}{2011}]{rincon2011}
{Herault} J.,  {Rincon} F.,  {Cossu} C.,  {Lesur} G.,  {Ogilvie} G.~I.,
  {Longaretti} P.-Y.,  2011, \mn@doi [\pre] {10.1103/PhysRevE.84.036321}, \href
  {http://adsabs.harvard.edu/abs/2011PhRvE..84c6321H} {84, 036321}

\bibitem[\protect\citeauthoryear{{Ji}, {Prager}  \& {Sarff}}{{Ji}
  et~al.}{1995}]{ji95}
{Ji} H.,  {Prager} S.~C.,   {Sarff} J.~S.,  1995, \mn@doi [Physical Review
  Letters] {10.1103/PhysRevLett.74.2945}, \href
  {http://adsabs.harvard.edu/abs/1995PhRvL..74.2945J} {74, 2945}

\bibitem[\protect\citeauthoryear{Kageyama, Ji, Goodman, Chen  \&
  Shoshan}{Kageyama et~al.}{2004}]{kageyama04}
Kageyama A.,  Ji H.,  Goodman J.,  Chen F.,   Shoshan E.,  2004, J. Phys. Soc.
  Japan, 73, 2424

\bibitem[\protect\citeauthoryear{{Keppens}, {Casse}  \& {Goedbloed}}{{Keppens}
  et~al.}{2002}]{keppens2002}
{Keppens} R.,  {Casse} F.,   {Goedbloed} J.~P.,  2002, \mn@doi [\apjl]
  {10.1086/340666}, \href {http://adsabs.harvard.edu/abs/2002ApJ...569L.121K}
  {569, L121}

\bibitem[\protect\citeauthoryear{{Lesur} \& {Ogilvie}}{{Lesur} \&
  {Ogilvie}}{2008}]{lesurdynamo}
{Lesur} G.,  {Ogilvie} G.~I.,  2008, \mn@doi [Astronomy \& Astrophysics]
  {10.1051/0004-6361:200810152}, \href
  {http://adsabs.harvard.edu/abs/2008A%26A...488..451L} {488, 451}

\bibitem[\protect\citeauthoryear{{Lesur} \& {Ogilvie}}{{Lesur} \&
  {Ogilvie}}{2010}]{Lesur2010}
{Lesur} G.,  {Ogilvie} G.~I.,  2010, \mn@doi [MNRAS]
  {10.1111/j.1745-3933.2010.00836.x}, \href
  {http://adsabs.harvard.edu/abs/2010MNRAS.404L..64L} {404, L64}

\bibitem[\protect\citeauthoryear{{Mitra} \& {Brandenburg}}{{Mitra} \&
  {Brandenburg}}{2012}]{mitra2012}
{Mitra} D.,  {Brandenburg} A.,  2012, \mn@doi [\mnras]
  {10.1111/j.1365-2966.2011.20190.x}, \href
  {http://adsabs.harvard.edu/abs/2012MNRAS.420.2170M} {420, 2170}

\bibitem[\protect\citeauthoryear{{Moffatt}}{{Moffatt}}{1978}]{moffat}
{Moffatt} H.~K.,  1978, {Magnetic field generation in electrically conducting
  fluids}

\bibitem[\protect\citeauthoryear{{Nauman} \& {Blackman}}{{Nauman} \&
  {Blackman}}{2014}]{Nauman2014}
{Nauman} F.,  {Blackman} E.~G.,  2014, \mn@doi [MNRAS] {10.1093/mnras/stu706},
  \href {http://adsabs.harvard.edu/abs/2014MNRAS.441.1855N} {441, 1855}

\bibitem[\protect\citeauthoryear{Noguchi, Pariev, Colgate  \& Nordhaus}{Noguchi
  et~al.}{2002}]{noguchi02}
Noguchi K.,  Pariev I.,  Colgate S.,   Nordhaus J.,  2002, Astrophys. J., 575,
  1151

\bibitem[\protect\citeauthoryear{{Ogilvie} \& {Pringle}}{{Ogilvie} \&
  {Pringle}}{1996}]{ogilvie96}
{Ogilvie} G.~I.,  {Pringle} J.~E.,  1996, \mn@doi [\mnras]
  {10.1093/mnras/279.1.152}, \href
  {http://adsabs.harvard.edu/abs/1996MNRAS.279..152O} {279, 152}

\bibitem[\protect\citeauthoryear{{Parker}}{{Parker}}{1979}]{Parker1979}
{Parker} E.~N.,  1979, {Cosmical magnetic fields: Their origin and their
  activity}.
Oxford Univ. Press

\bibitem[\protect\citeauthoryear{{Regev} \& {Umurhan}}{{Regev} \&
  {Umurhan}}{2008}]{regev08}
{Regev} O.,  {Umurhan} O.~M.,  2008, \mn@doi [Astronomy \& Astrophysics]
  {10.1051/0004-6361:20078413}, \href
  {http://adsabs.harvard.edu/abs/2008A%26A...481...21R} {481, 21}

\bibitem[\protect\citeauthoryear{R\"udiger, Schultz  \& Shalybkov}{R\"udiger
  et~al.}{2003}]{rudiger03}
R\"udiger G.,  Schultz M.,   Shalybkov D.,  2003, Phys. Rev. E, 67, 046312

\bibitem[\protect\citeauthoryear{Schnack, Barnes, Mikic, Harned  \&
  Caramana}{Schnack et~al.}{1987}]{Debs}
Schnack D.~D.,  Barnes D.~C.,  Mikic Z.,  Harned D.~S.,   Caramana E.~J.,
  1987, Journal of Computational Physics, 70, 330

\bibitem[\protect\citeauthoryear{{Simon}, {Hawley}  \& {Beckwith}}{{Simon}
  et~al.}{2011}]{simon2011}
{Simon} J.~B.,  {Hawley} J.~F.,   {Beckwith} K.,  2011, \mn@doi [\apj]
  {10.1088/0004-637X/730/2/94}, \href
  {http://adsabs.harvard.edu/abs/2011ApJ...730...94S} {730, 94}

\bibitem[\protect\citeauthoryear{Sisan, Mujica, Tillotson, Huang, Dorland,
  Hassam, Antonsen  \& Lathrop}{Sisan et~al.}{2004}]{sisan04}
Sisan D.,  Mujica N.,  Tillotson W.,  Huang Y.-M.,  Dorland W.,  Hassam A.,
  Antonsen T.,   Lathrop D.,  2004, Phys. Rev. Lett., 93, 114502

\bibitem[\protect\citeauthoryear{{Sorathia}, {Reynolds}, {Stone}  \&
  {Beckwith}}{{Sorathia} et~al.}{2012}]{Sorathia2012}
{Sorathia} K.~A.,  {Reynolds} C.~S.,  {Stone} J.~M.,   {Beckwith} K.,  2012,
  \mn@doi [\apj] {10.1088/0004-637X/749/2/189}, \href
  {http://adsabs.harvard.edu/abs/2012ApJ...749..189S} {749, 189}

\bibitem[\protect\citeauthoryear{{Squire} \& {Bhattacharjee}}{{Squire} \&
  {Bhattacharjee}}{2015}]{Squire2015b}
{Squire} J.,  {Bhattacharjee} A.,  2015, preprint, \href
  {http://adsabs.harvard.edu/abs/2015arXiv150801566S} {} (\mn@eprint {arXiv}
  {1508.01566})

\bibitem[\protect\citeauthoryear{{Sridhar} \& {Singh}}{{Sridhar} \&
  {Singh}}{2014}]{sridhar2014}
{Sridhar} S.,  {Singh} N.~K.,  2014, \mn@doi [\mnras] {10.1093/mnras/stu1981},
  \href {http://adsabs.harvard.edu/abs/2014MNRAS.445.3770S} {445, 3770}

\bibitem[\protect\citeauthoryear{Stefani, Gundrum, Gerbeth, R\"udiger, nd
  J.~Szklarski  \& Hollerbach}{Stefani et~al.}{2007}]{tefani07}
Stefani F.,  Gundrum T.,  Gerbeth G.,  R\"udiger G.,  nd J.~Szklarski M.~S.,
  Hollerbach R.,  2007, Phys. Rev. Lett., 97, 184502

\bibitem[\protect\citeauthoryear{{Suzuki} \& {Inutsuka}}{{Suzuki} \&
  {Inutsuka}}{2014}]{Suzuki2014}
{Suzuki} T.~K.,  {Inutsuka} S.-i.,  2014, \mn@doi [\apj]
  {10.1088/0004-637X/784/2/121}, \href
  {http://adsabs.harvard.edu/abs/2014ApJ...784..121S} {784, 121}

\bibitem[\protect\citeauthoryear{Velikhov}{Velikhov}{1959}]{velikhov59}
Velikhov E.~P.,  1959, Sov. Physics JETP, 36, 995

\bibitem[\protect\citeauthoryear{{Vishniac} \& {Brandenburg}}{{Vishniac} \&
  {Brandenburg}}{1997}]{vishniac97}
{Vishniac} E.~T.,  {Brandenburg} A.,  1997, \apj, \href
  {http://adsabs.harvard.edu/abs/1997ApJ...475..263V} {475, 263}

\bibitem[\protect\citeauthoryear{Yousef \& et al.}{Yousef \&
  et~al.}{2008}]{Yousef2008}
Yousef T.~A.,  et al. 2008, \mn@doi [Phys. Rev. Lett.]
  {10.1103/PhysRevLett.100.184501}, \href
  {http://adsabs.harvard.edu/abs/2008PhRvL.100r4501Y} {100, 184501}

\makeatother
\end{thebibliography}
\end{document}